\documentclass[headings=standardclasses]{scrartcl}

\usepackage{geometry}
\usepackage[utf8]{inputenc}
\usepackage[T1]{fontenc}
\usepackage[charter]{mathdesign}
\usepackage[onehalfspacing]{setspace}
\usepackage{microtype}
\usepackage{amsmath}
\usepackage{amsthm}
\usepackage{tabulary}
\usepackage{longtable}
\usepackage{booktabs} 
\newcommand{\ra}[1]{\renewcommand{\arraystretch}{#1}}
\usepackage{siunitx}
\usepackage{bm}
\usepackage[pdftex]{graphicx}
\usepackage{rotating}
\usepackage{pdfpages}
\usepackage{pdflscape}
\usepackage{color}
\usepackage{array}
\usepackage[font=bf]{caption}
\usepackage{pgfplots}
\usepackage{pgfplotstable}
\usepackage{float}
\usepackage{appendix}
\pgfplotsset{compat=1.9}
\usepackage{url}
\usepackage[section]{placeins}
\usepackage{natbib}
\usepackage{hyperref}
\hypersetup{
	colorlinks   = true,    % Colours links instead of ugly boxes
	urlcolor     = blue,    % Colour for external hyperlinks
	linkcolor    = blue,    % Colour of internal links
	citecolor    = blue      % Colour of citations
}

\begin{document}
\begin{spacing}{1.2}
\begin{flushleft}
\huge \textbf{Autonomous Driving and Residential Location Preferences:
Evidence from a Stated Choice Survey} \\
\bigskip 
\normalsize
20 September 2019 \\
\vspace{\baselineskip}
Rico Krueger\\
Research Centre for Integrated Transport Innovation, School of Civil and Environmental Engineering, UNSW Australia, Sydney NSW 2052, Australia \\
r.krueger@student.unsw.edu.au \\
\vspace{\baselineskip}
Taha H. Rashidi (corresponding author)  \\
Research Centre for Integrated Transport Innovation, School of Civil and Environmental Engineering, UNSW Australia, Sydney NSW 2052, Australia\\
rashidi@unsw.edu.au\\
\vspace{\baselineskip}
Vinayak V. Dixit \\
Research Centre for Integrated Transport Innovation, School of Civil and Environmental Engineering, UNSW Australia, Sydney NSW 2052, Australia\\
rashidi@unsw.edu.au\\
\end{flushleft}
\end{spacing}

\newpage
\section*{Abstract}

%Motivation
The literature suggests that autonomous vehicles (AVs) may drastically change the user experience of private automobile travel by allowing users to engage in productive or relaxing activities while travelling. As a consequence, the generalised cost of car travel may decrease, and car users may become less sensitive to travel time. 
%Problem
By facilitating private motorised mobility, AVs may eventually impact land use and households' residential location choices. 
%Method
This paper seeks to advance the understanding of the potential impacts of AVs on travel behaviour and land use by investigating stated preferences for combinations of residential locations and travel options for the commute in the context of autonomous automobile travel. Our analysis draws from a stated preference survey, which was completed by 512 commuters from the Sydney metropolitan area in Australia and provides insights into travel time valuations in a long-term decision-making context. For the analysis of the stated choice data, mixed logit models are estimated. 
%Results
Based on the empirical results, no changes in the valuation of travel time due to the advent of AVs should be expected. However, given the hypothetical nature of the stated preference survey, the results may be affected by methodological limitations.

\vspace{\baselineskip}
\noindent \emph{Keywords:} connected and autonomous vehicles, residential location choice, value of time, stated choice, mixed logit

\newpage

\section{Introduction}

%Motivation
In light of recent technological advancements in the fields of vehicle automation and connection, connected and autonomous vehicles (AVs) with advanced self-driving capabilities are anticipated to be available to the consumer mass market in the not-so-distant future \citep[e.g.][]{burns2013vision, fagnant2015preparing, nieuwenhuijsen2018towards, wadud2016help}. In their technologically most advanced stage, AVs are expected to be fully automated and to operate autonomously, not requiring any control input by the user \citep{nhtsa2013preliminary, saei2016taxonomy}. AVs may catalyse a profound transformation of the transport sector by making private motorised mobility safer \citep[e.g.][]{anderson2014autonomous, fagnant2015preparing, sparrow2017when}, more affordable \citep[e.g.][]{bosch2018cost-based, burns2013transforming} and more accessible to user groups that are currently unwilling or unable to drive \citep[e.g.][]{anderson2014autonomous, fagnant2015preparing, harper2016estimating}. Moreover, AVs with full self-driving capabilities may substantially alter the user experience of private motorised mobility by allowing users to engage in productive or relaxing activities while travelling in an AV \citep[e.g.][]{le_vine2015autonomous, pudane2018time}. In turn, some travellers' generalised cost of car travel may decrease, and a proportion of travellers may become less sensitive to travel time. As AVs facilitate private motorised mobility, the accessibility of locations may increase and eventually, AVs may fuel urban sprawl and suburbanisation by affecting the location choices of households and firms \citep{meyer2017autonomous, milakis2018implications}. 

%Problem and background
Strategic policy processes in the domain of integrated transport and land use planning depend on future-proof estimates of population sensitivities with respect to travel- and lifestyle-related variables \citep[e.g.][]{bhat2007comprehensive, pinjari2007modeling, schwanen2005what}. The lifestyle-based approach to travel behaviour analysis recognises complex interdependencies between travel behaviour, residential location and lifestyle orientations \citep{van_acker2010transport}: It is understood that travel behaviour (e.g. mode choice) and residential location are co-determined by lifestyle goals pertaining to travel, neighbourhood and housing \citep{bhat2007comprehensive, pinjari2007modeling, schwanen2005what}. As a consequence, a household's housing search process involves trade-offs between a variety of travel- and lifestyle-related variables \citep{guo2018modeling, rouwendal2001preferences, walker2006latent}; in particular, the spatial search process requires households to balance housing expenses against household members' travel times and costs to reach frequently-visited destinations such as the workplace \citep{so2001effects}. With the advent of AVs, some travellers' generalised cost of car travel may decrease. From this supposition, the question arises as to what extent the advent of AVs may affect preferences for travel and residential location.

%Objective and method
The current paper seeks to advance the understanding of the potential impacts of autonomous driving on travel behaviour and land use by investigating preferences for residential locations and travel options for the commute in the context of autonomous driving. To this end, we designed and implemented a stated preference survey, which was completed by 512 commuters from the Sydney metropolitan area in Australia. The survey featured a discrete choice experiment requiring respondents to jointly choose a housing option and a travel option for the commute, whereby the presented travel options included the transport modes conventional car, self-driving car and public transit. In multiple scenarios, attributes such as housing and commute costs as well as commute travel times were manipulated. For the analysis of the stated choice data, a mixed multinomial logit model accommodating unobserved taste variation and flexible substitution patterns is estimated. The estimation results provide insights into travel time valuations for autonomous driving in a long-term decision-making context.

%Paper outline
We organise the remainder of this paper as follows: First, we present a review of pertinent literature. Then, we describe the survey design and the data collection and explain the modelling approach. Next, we present the results. We conclude with a general discussion of the findings. 

\section{Literature review}

A rapidly-expanding body of literature is concerned with evaluating the economic, societal and environmental implications of autonomous driving \citep[for a comprehensive review, see][]{milakis2017policy}. Within this body of literature, several studies explicitly investigate the impacts of autonomous driving on travel behaviour and land use. \citet{soteropoulos2018impacts} present a systematic review of 37 studies on the topic. The majority of these studies rely on activity-based or agent-based modelling approaches and find that the advent of AVs may lead to increases in vehicle kilometres travelled, lower levels of public transit and active mode use as well as to suburbanisation and urban sprawl. \citet{soteropoulos2018impacts} observe that the findings of the considered studies may be sensitive to the underlying modelling assumptions, particularly in regards to the value of time for autonomous driving and the business models under which AVs become available. 

Of the studies reviewed by \citet{soteropoulos2018impacts}, we highlight three studies that focus on residential location choices. Using a spatial general equilibrium model, \citet{gelauff2017spatial} analyse changes in home and job location as well as in commute mode choice in the context of two automation scenarios in the Netherlands. In the first scenario, AVs permit productive time use during long-distance car travel; in the second scenario, AVs enable efficient door-to-door transport. The authors find that the first scenario results in urban sprawl, while the second scenario results in population concentration in core area. Employing a strategic land use transport interaction model, \citet{thakur2016urban} investigate changes in residential location choices and travel behaviour in response to the availability of AVs in the Melbourne metropolitan area in Australia. Assuming that the value of time for travel by AV is 50\% of the value of time for travel by conventional car, the authors forecast a population reduction of 4\% in central areas and a population increase of 3\% in outer suburbs. \citet{zhang2018residential} predict residential location choices in the Atlanta metropolitan area in the USA in the presence of shared autonomous vehicles \citep[SAVs, e.g.][]{fagnant2014travel, krueger2016preferences}. The study relies on an agent-based simulation model, which couples a residential location choice model assuming extant residential location preferences and a SAV simulation model assuming that the value of time for travel by SAV is between 0\% and 75\% of the current value of time for travel by conventional car. The authors find that the presence of SAVs may result in an increase in commute vehicle kilometres travelled across all considered market segments. 

Analysing data sourced from a survey among 347 individuals from the city of Austin in the USA, \citet{bansal2016assessing} develop an ordered probit model to explain whether respondents would shift their home locations, if AVs and SAVs became available. The authors find that respondents who live in dense residential neighbourhoods, who have a greater number of children and who have obtained a Bachelor's degree are comparatively more likely to indicate that they would consider to live further away from central areas; by contrast, respondents who are familiar with car sharing and who carry a smartphone are comparatively more likely to indicate that they would consider to move closer to central areas, if AVs and SAVs became available. Relatedly, \citet{bansal2018are} draw from a questionnaire-based survey in Texas, USA and find that 81.5\% of 1,088 respondents would not consider to change their home location, once AVs become available. Moreover, \citet{milakis2018implications} apply the Q-method to analyse expert opinions on the possible accessibility impacts of autonomous driving. The authors extract three main viewpoints from 17 expert responses: First, initial accessibility gains may be neutralised by increased travel demand; second, autonomous driving will lead to increased density in the urban core and to decreased density in surrounding areas; third; the accessibility benefits will be unequally distributed and will only be enjoyed by those who can afford autonomous driving.

A second growing body of literature investigates preferences for different aspects of autonomous driving with the help of discrete choice experiments \citep[see][for a comprehensive review]{gkartzonikas2019have}. The studies within this body of literature consider short- and long-term decision contexts, but interdependencies between preferences for residential location and travel behaviour have not been explicitly investigated. \citet{daziano2017are} analyse stated preferences of households in the USA for different levels of private vehicle automation. Relatedly, \citep{shabanpour2018eliciting} investigate stated preferences of individuals living in the Chicago metropolitan area in the USA for attributes and ownership of AVs. Moreover, \citet{haboucha2017user} examine stated preferences of commuters in Israel and the USA for owning and sharing AVs. \citet{krueger2016preferences} and \citet{winter2017stated} examine stated preferences for the use of SAVs in Australia and respectively the Netherlands. Similarly, \citet{kolarova2017estimation} investigate stated preferences for the use of SAVs and privately-owned AVs in Germany.

Preferences for combinations of housing and travel options for the commute have been studied in a general context without consideration of autonomous driving. For example, \citet{rouwendal2001preferences} analyse Dutch workers' stated preferences for combinations of housing, employment and commuting, using conditional and mixed multinomial logit models. The authors find that workers are generally averse to commuting but may accept longer commutes in exchange for some housing attributes. In particular, dual-earner households are found to show a preference for living in small- and medium-sized cities. \citet{walker2006latent} estimate a latent class multinomial logit model to analyse stated lifestyle and travel preferences of residents of Portland, Oregon, USA. The authors discern three lifestyle segments, namely suburban dwellers who prefer auto-oriented lifestyles with larger residences in suburban neighbourhoods that offer services and amenities nearby, transit riders who prefer to public-transportation-oriented lifestyles in suburban neighbourhoods, and urban dwellers who prefer auto-oriented lifestyles in smaller homes in urban neighbourhoods. Using stated choice data collected in Shenyang, China, \citet{guo2018modeling} develop a mixed multinomial logit model with error components to jointly analyse preferences for home relocation, job change and commute mode choice. The authors find substantial unobserved heterogeneity in tastes and substitution patterns. 

To conclude, a review of the pertinent literature reveals that the potential impacts of autonomous driving on trade-offs between key travel- and lifestyle-related variables have not been sufficiently quantified. The current paper seeks to advance the literature in this regard by investigating stated preferences for combinations of residential location options and different commute travel options in the context of autonomous driving. 

\section{Methodology}

\subsection{Survey design and data collection}

Data for this research study were collected via a web-based stated preference survey, which was completed by 512 residents of the Sydney metropolitan area in Australia. The survey comprised two parts: The first part requested information about respondents' socio-demographic characteristics and their current commuting and housing arrangements; the second part featured a stated choice experiment requiring respondents to jointly choose a housing option and a travel option for the commute.

\subsubsection{Choice tasks and alternatives}
\begin{figure}[p]
\centering
\includegraphics[width = 0.85 \textwidth]{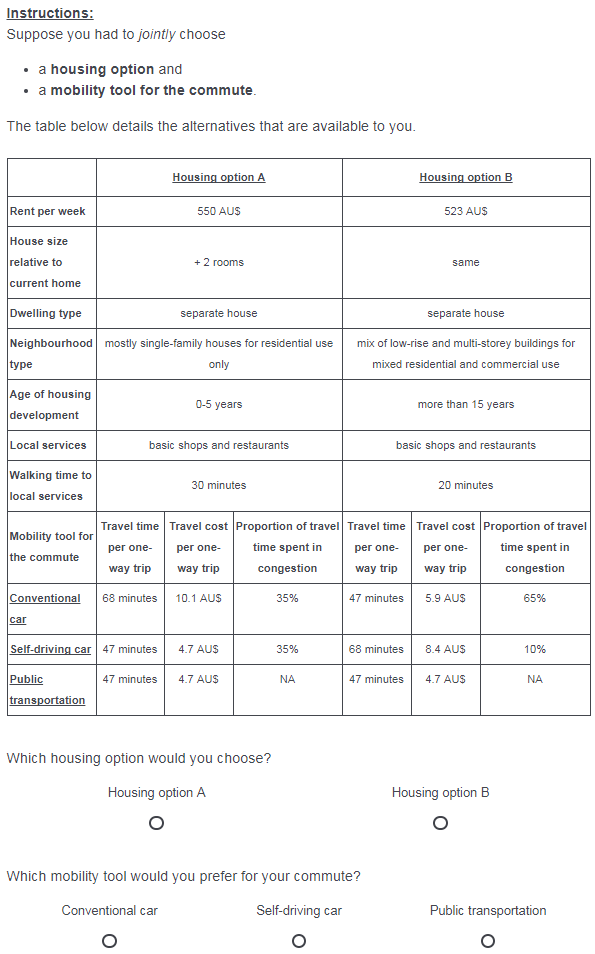}
\caption{Example of a stated choice task} \label{f_conjoint}
\end{figure}

The stated choice experiment consisted of eight choice tasks, which required respondents to jointly choose one of two housing options and a travel option for the commute. The travel options were labelled ``conventional car'', ``self-driving car'' and ``public transportation''; if a respondent indicated that she did not hold a valid driving license, the option labelled ``conventional car'' was not displayed. The choice tasks were presented in a conjoint format (see Figure \ref{f_conjoint}).

\FloatBarrier
\subsubsection{Attributes}
The housing options were characterised by seven attributes, namely the weekly housing cost, the house size relative to the respondent's current home, the dwelling type, the age of the housing development, the neighbourhood type, the presence of local services and the walking time to local services. If a respondent indicated to be a renter, the attribute was labelled ``rent per week''. If a respondent indicated to live in a mortgaged home, the attribute was labelled ``mortgage payment per week''. The selection of the housing attributes was inspired by other stated choice experiments that have appeared in the literature \citep{dubernet2016choice, rouwendal2001preferences, walker2006latent}. The travel options for the commute were characterised by three attributes, namely the travel time per one-way trip, the travel cost per one-way trip and the proportion of travel time usually spent in congestion.\footnote{Here we choose to display travel costs on a per-trip basis rather than on a per week or a per month basis for two reasons. First, we wish to display travel times and cost in a consistent manner within and across alternatives. Second, public transit fares are calculated on a per trip basis largely independent of usage intensity and season ticket are not available; therefore, respondents are most familiar with travel costs shown on a per trip basis.} Each of the housing options entailed different attribute values for the commute travel options such that the two choice dimensions depended on one another. To personalise the choice tasks, the attributes housing cost and travel time were pivoted around the respondent's status quo. Table \ref{table_design} gives a summary of the attributes and their levels. The experimental design of the choice tasks is based on orthogonal main-effect arrays and was created with the support.CE package for R \citep{aizaki2012basic}.

\begin{table}[H]
\small
\centering
\ra{1.2}
\begin{tabular}{@{}p{4cm} p{10cm}@{}}
\toprule
\textbf{Attribute}  & \textbf{Description and levels} \\
\midrule

\textbf{Housing} \\

Housing cost & 
\begin{minipage}[t]{0.6\textwidth}
    \begin{itemize}
    \item Three levels, pivoted around status quo
    	\begin{itemize}
	\item reference = $\min \{$ 6,500 AUD, $\max \{$150 AUD, current weekly housing cost in AUD$\} \}$
	\item pivot factors = $\{$0.95, 1.00, 1.05$\}$ 
	\end{itemize}
    \end{itemize}
\end{minipage}
\\

Additional number of rooms &
\begin{minipage}[t]{0.6\textwidth}
    \begin{itemize}
    \item Three levels, not pivoted
    	\begin{itemize}
	\item levels = $\{$0, 1, 2$\}$
	\end{itemize}
    \end{itemize}
\end{minipage}
\\

Dwelling type &
\begin{minipage}[t]{0.6\textwidth}
    \begin{itemize}
    \item Three levels, not pivoted
    	\begin{itemize}
	\item levels = $\{$unit, townhouse, separate house$\}$ 
	\end{itemize}
    \end{itemize}
\end{minipage}
\\

Neighbourhood type &
\begin{minipage}[t]{0.6\textwidth}
    \begin{itemize}
    \item Three levels, not pivoted
    	\begin{itemize}
	\item levels = $\{$unit, townhouse, separate house$\}$
	\end{itemize}
    \end{itemize}
\end{minipage}
\\
		                 
Local services &
\begin{minipage}[t]{0.6\textwidth}
    \begin{itemize}
    \item Three levels, not pivoted
    	\begin{itemize}
	\item levels = $\{$no shops, basic shops and restaurants, basic plus specialty shops and restaurants$\}$
	\end{itemize}
    \end{itemize}
\end{minipage}
\\
		                 
Walking time to local services, if local services are available &
\begin{minipage}[t]{0.6\textwidth}
    \begin{itemize}
    \item Three levels, not pivoted
    	\begin{itemize}
	\item levels = $\{$10 minutes, 20 minutes, 30 minutes$\}$ 
	\end{itemize}
    \end{itemize}
\end{minipage}
\\
		
\textbf{Mobility tools} \\	

Travel time &
\begin{minipage}[t]{0.6\textwidth}
    \begin{itemize}
    \item Three levels, pivoted around status quo
    	\begin{itemize}
	\item reference = $\min \{$80 minutes, $\max \{$30 minutes, current commute time per one-way-trip in minutes$\} \}$
	\item pivot factors = $\{$1.05, 1.25, 1.5$\}$
	\end{itemize}
    \end{itemize}
\end{minipage}
\\        	                 	                 

Travel cost &
\begin{minipage}[t]{0.6\textwidth}
    \begin{itemize}
    \item Three levels, pivoted
    	\begin{itemize}
	\item reference = hypothetical travel time of the mode
	\item pivot factors = $\{$0.100 $\frac{\mbox{AUD}}{\mbox{minutes}}$, 0.125 $\frac{\mbox{AUD}}{\mbox{minutes}}$, 0.150 $\frac{\mbox{AUD}}{\mbox{minutes}}$$\}$
	\end{itemize}
    \end{itemize}
\end{minipage}
\\ 

Proportion of travel time spent in congestion &
\begin{minipage}[t]{0.6\textwidth}
    \begin{itemize}
    \item Three levels, not pivoted
    	\begin{itemize}
	\item levels = $\{$10\%, 35\%, 65\%$\}$ 
	\end{itemize}
    \end{itemize}
\end{minipage}
\\

\bottomrule
\end{tabular}
\caption{Attributes and attribute levels} \label{table_design}
\end{table}

\subsubsection{Instructions}
Prior to the presentation of the choice tasks, respondents were shown detailed instructions about the stated choice experiment. Respondents were told that they had to jointly choose a housing option and a preferred travel option for the commute. Moreover, the nature of the the travel options was explained to the respondents. Images were used to illustrate the experience of using one of the travel options. The travel option ``self-driving car'' was defined in accordance with SAE level 5 \citep{saei2016taxonomy}: It was highlighted that a driving license would not be required to operate the vehicle, as the vehicle would be fully self-driving and would not require any input by the user. Furthermore, it was emphasised that users could engage in productive or relaxing activities while travelling in a self-driving car.

\subsubsection{Implementation and data collection}
The stated choice survey was implemented using the Qualtrics software platform and distributed to an Australian consumer panel in July 2017. The target population was the general adult population of the Sydney metropolitan area in Australia but given the study objective, several inclusion criteria were imposed on the study participants: First, respondents had to be employed or studying; second, respondents had to commute at least three times per week; third, respondents had to be renting a home or owning a home with a mortgage. In addition, quotas on the commute duration, the tenure type and the current housing costs were imposed to increase the socio-economic diversity of the sample. Despite these efforts, the representativity of the sample with respect to the general adult population of the Sydney metropolitan cannot be guaranteed given the employed recruitment method. We refer to census data for more information about the characteristics of the target population \citep[see][]{abs2019}. 

\subsection{Modelling approach} \label{S_modelling}

For the analysis of the stated choice data, we develop a mixed multinomial logit (M-MNL) model \citep{train2009discrete}, which accommodates unobserved taste variation \citep{mcfadden2000mixed} and flexible substitution patterns \citep{brownstone1999forecasting}. The model is established as follows: Decision-makers are assumed to be utility maximisers. In choice scenario $t \in \{1, \ldots T_{n} \}$, decision-maker $n \in \{1, \ldots N \}$ derives utility 
\begin{equation}
U_{n,t,j} = V(\boldsymbol{X}_{n,t,j}, \boldsymbol{\beta}_{n}) + \epsilon_{n,t,j}
\end{equation}
from alternative $j \in C_{n,t}$. Here $V()$ denotes the representative utility, $\boldsymbol{X}_{n,t,j}$ is a row-vector of covariates, $\boldsymbol{\beta}_{n}$ is a collection of taste parameters, and $\epsilon_{n,t,j}$ is a stochastic disturbance. The assumption $\epsilon_{n,t,j} \sim \mbox{Gumbel}(0,1)$ leads to the multinomial logit model such that the probability that decision-maker $n$ chooses alternative $j \in C_{n,t}$ in scenario $t$ is given by
\begin{equation} \label{eq_prob}
P(y_{n,t} = j \vert \boldsymbol{X}_{n,t},  \boldsymbol{\beta}_{n}) = \frac{\exp \left ( V(\boldsymbol{X}_{n,t,j}, \boldsymbol{\beta}_{n}) \right )}{\sum_{j' \in C_{n,t}} \exp \left ( V(\boldsymbol{X}_{n,t,j'}, \boldsymbol{\beta}_{n}) \right )},
\end{equation}
where $y_{n,t} \in C_{n,t}$ captures the observed choice. The choice probability can be iterated over choice scenarios to obtain the probability of observing a decision-maker's sequence of choices $\boldsymbol{y}_{n}$: 
\begin{equation} \label{eq_condProb}
P(\boldsymbol{y}_{n} \vert \boldsymbol{X}_{n},  \boldsymbol{\beta}_{n}) = \prod_{t = 1}^{T_{n}} P(y_{n,t} = j \vert \boldsymbol{X}_{n,t},  \boldsymbol{\beta}_{n}).
\end{equation}
$\boldsymbol{\beta}_{n}$ are assumed to be realisations from some distribution D described by parameter $\boldsymbol{\theta}$, i.e. $\boldsymbol{\beta}_{n} \sim D(\boldsymbol{\theta})$,  $n = 1, \ldots, N$, whereby $f$ denotes the density of $D$. To account for the fact that a decision-maker's tastes are unobserved, we marginalise (\ref{eq_condProb}) over the taste parameter distribution to obtain the unconditional probability of a decision-maker's sequence of choices: 
\begin{equation}
P(\boldsymbol{y}_{n} \vert \boldsymbol{X}_{n},  \boldsymbol{\theta}) = \int P(\boldsymbol{y}_{n} \vert \boldsymbol{X}_{n},  \boldsymbol{\beta}_{n}) f(\boldsymbol{\beta}_{n} \vert \boldsymbol{\theta}) d \boldsymbol{\beta}_{n}
\end{equation}
Iterating this unconditional probability over all decision-makers in the sample yields the likelihood function of the M-MNL model:
\begin{equation}
\mathcal{L}(\boldsymbol{y} \vert \boldsymbol{X}, \boldsymbol{\theta}) = \prod_{n = 1}^{N} P(\boldsymbol{y}_{n} \vert \boldsymbol{X}_{n},  \boldsymbol{\theta}).
\end{equation}

The present application requires us to recognise that the dependent variable considered in our analysis is a choice of a combination of a housing option and a travel option for the commute. In other words, each $j$ is a tuple $(k, l)$, where $k \in \{1, 2 \}$ indexes the two unlabelled housing options and where $l \in \{1, 2, 3 \}$ indexes the three transport modes conventional car, self-driving car and public transportation. The utility function for each tuple $j = (k, l)$ is then specified as
\begin{equation}
U_{n,t,(k, l)} = \boldsymbol{X}_{n,t,k} \boldsymbol{\beta}_{H, n} + \boldsymbol{X}_{n,t,l} \boldsymbol{\beta}_{M, n} + \eta_{n,l} + \epsilon_{n,t,(k, l)}
\end{equation}
where $\boldsymbol{\beta}_{n}$ is partitioned into random taste parameters $\boldsymbol{\beta}_{H, n}$ pertaining to the attributes of the housing options and random taste parameters $\boldsymbol{\beta}_{M, n}$ pertaining to the attributes of the travel options. In the present application, each element $\beta_{n,r}$ of $\boldsymbol{\beta}_{n}$ is given by a weakly monotonic transformation $\psi_{r}$ of an auxiliary normal random parameter $\alpha_{n,r}$ \citep[see][]{train2005mixed}, i.e. $\beta_{n,r} = \psi_{r}( \alpha_{n,r} )$ for $n = 1, \dots, N$ and $r = 1, \dots, R$, whereby $\boldsymbol{\alpha}_{n} \sim \mbox{N}(\boldsymbol{\mu}, \boldsymbol{\Sigma})$ for $n = 1, \dots, N$. The functional form of the transformation $\psi_{r}$ determines the distribution of $\beta_{1:N,r}$: 
The identity function defines normally distributed tastes, and exponentiation results in log-normally distributed tastes.
%, censoring allows for truncated normally distributed tastes, and the logistic transformation leads to the $S_{B}$ distribution. 
Moreover, $\eta_{n,l}$ are normal error components, which allow for flexible substitution patterns by inducing correlation across the utilities of alternatives within three nests of transport modes, $\eta_{n,l} \sim \mbox{N}(0, \tau_{\eta_{l}}^{2})$, $n = 1, \ldots, N$, $l = 1, 2, 3$, where $\tau_{\eta_{l}}$, $l = 1, 2, 3$ are scale parameters. Since we rely on panel data for model estimation, the scale parameters of all heteroskedastic error components are identified \citep[see][]{walker2007identification}. 

For model estimation, we employ maximum simulated likelihood methods \citep[e.g.][]{train2009discrete}, which we implement by writing our own MATLAB code. Choice probabilities are simulated using 1,024 points from a scrambled and shifted Sobol' quasi-random sequence \citep{matousek1998on,sobol1967on}.

\section{Results}

\subsection{Sample description}

Valid responses were collected from 512 individuals. Relative frequencies of the respondents' socio-economic characteristics are reported in Table \ref{table_dataResp}. As each respondent completed eight choice tasks, a total of 4,096 cases were observed. The travel option self-driving car was selected in 27.2\% of the cases; the travel options conventional car and public transportation were selected in 35.4\% and 37.3\% of the cases, respectively. 56.6\% of the respondents selected the travel option self-driving car at least once. Table \ref{table_dataChosen} presents a description of the observed choices in terms of the relative attribute levels of the chosen and non-chosen alternatives for selected attributes and alternatives. 

Additional descriptive statistics on the collected stated choice data are reported in Tables \ref{table_dataChosenHousing} and \ref{table_dataChosenMob} in the appendix.

\begin{table}[H]
\small
\centering
\ra{1.2}
\begin{tabular}{@{}l p{4cm} l @{}}
\toprule
\textbf{Characteristic} & \textbf{Values}   & \textbf{Sample distribution} \\
\midrule
\textbf{Socio-demographic characteristics} \\
Age & [18--29 years old, 30--49 years old, 50 -- 64 years old, 65 years old or older] & [17.2\%, 53.3\%, 24.6\%, 3.7\%] \\
Gender  & [female, male, other] & [47.1\%, 52.5\%, 0.4\%]\\
Weekly household income [AUD] & [0--799, 800--1,599, 1,600--2,499, 2,500 or more] & [14.9\%, 28.8\%, 31.2\%, 25.2\%] \\
Highest level of education completed & [post-secondary or lower, Bachelors degree or higher] & [43.4\%, 56.6\%] \\
Presence of children in household & [yes, no] & [33.8\%, 66.2\%]  \\
\textbf{Mobility attributes} \\
Holds driving license & [yes, no] & [91.4\%, 8.6\%]\\
Ride-sharing service user & [yes, no] & [58.4\%, 41.6\%]\\
\textbf{Housing arrangements} \\
Tenure type & [renting, owning with mortgage] & [44.9\%, 55.1\%] \\
Weekly housing cost [AUD] & [299 or less, 300--429, 430--599, 600 or more] & [25.0\%, 25.0\%,  25.0\%, 25.0\%] \\
Dwelling type & [separate house, townhouse, unit] & [45.9\%, 15.0\%, 39.1\%] \\
Home layout & [studio,1--2 bedrooms, 3 bedrooms, 4 bedrooms, 5 bedrooms or more] & [8.4\%, 13.6\%, 27.1\%, 27.5\%, 23.3\%] \\
Neighbourhood type & [city centre, inner suburb, outer suburb, rural] & [14.6\%, 42.2\%, 41.2\%, 2.0\%] \\
\textbf{Commute} \\
Main commute mode & [car as driver, car as passenger, public transportation, bicycle, walk] & [55.1\%, 3.7\%, 35.0\%, 1.4\%, 4.9\%] \\
Duration (one-way) [min] & [19 or less, 20--29, 30--49,50--150] & [25.0\%, 25.0\%, 25.0\%, 25.0\%] \\
\bottomrule
\end{tabular}
\caption{Summary statistics of characteristics of the respondents (N = 512)} \label{table_dataResp}
\end{table}

\begin{table}[h]
\small
\centering
\ra{1.2}
\begin{tabular}{@{}p{12cm} c c@{}}
\toprule
& \textbf{Rel.} & \\ 
\textbf{Condition and event}  & \textbf{freq.} & \textbf{N} \\
\midrule

Given the choice of self-driving car as a mobility option for the commute \ldots \\
\quad \ldots how frequently was self-driving car selected even though there was a faster travel option available for the same housing option? & 0.41 & 1116\\
\quad \ldots how frequently was self-driving car selected even though conventional car was faster for the same housing option? & 0.26 & 1116 \\
\quad \ldots how frequently was self-driving car selected even though public transit was faster for the same housing option? & 0.26 & 1116\\

Given the choice of self-driving car as a mobility option for the commute \ldots \\
\quad \ldots how frequently was self-driving car selected even though there was a less expensive travel option available for the same housing option? & 0.39 & 1116\\
\quad \ldots how frequently was self-driving car selected even though conventional car was less expensive for the same housing option? & 0.27 & 1116\\
\quad \ldots how frequently was self-driving car selected even though public transit was less expensive for the same housing option? & 0.26 & 1116\\

Given the choice of conventional car as a mobility option for the commute \ldots \\
\quad \ldots how frequently was conventional car selected in combination with the less expensive housing option? & 0.41 & 1452 \\
Given the choice of self-driving car as a mobility option for the commute \ldots \\
\quad \ldots how frequently was self-driving car selected in combination with the less expensive housing option? & 0.36 & 1116 \\
Given the choice of public transit as a mobility option for the commute \ldots \\
\quad \ldots how frequently was public transit selected in combination with the less expensive housing option? & 0.43 & 1528 \\

\bottomrule
\end{tabular}
\caption{Description of observed choices} \label{table_dataChosen}
\end{table}

\FloatBarrier
\subsection{Final model specification}

The final model specification is the product of an extensive specification search. In the final model specification, sensitivities to some attributes are treated as fixed parameters in the interest of parsimony or for better interpretability. In particular, sensitivities to costs are treated as fixed parameters to facilitate the derivation of willingness to pay distributions. To assure that the cost sensitivities are strictly negative and to assure that the standard errors of the parameters of the willingness to pay distributions are defined, the respective taste parameters are exponentiated and multiplied by negative one prior to being entered into the utility function \citep{carson2018new}. We allow also for systematic heterogeneity in the sensitivity to housing cost by estimating one taste parameter for owners and another one for renters; in addition, we follow the example of \citet{bhat2018new} and divide the respective attribute values for each respondent by the respondent's weekly household income so that the sensitivities to housing cost become functions of weekly household income.\footnote{In the questionnaire accompanying the stated choice experiment, income information was surveyed using twelve income bands. The ordinal data are transformed into continuous data by taking the midpoint of each response category; the highest response category (``4,000 AUD/week or more'') is top-coded to $1.3 \cdot 4,000$ AUD/week = 5,200 AUD/week.}

All random taste parameters follow a normal distribution, with exception of the sensitivity to congested travel conditions. To assure that sensitivities to congested travel conditions are strictly negative, the taste parameter corresponding to the negative of the attribute is assumed to follow a log-normal distribution.\footnote{We also explored estimating mode-specific sensitivities to congestion but found that the differences in sensitivities were not statistically significant. In the interest of parsimony, the final model specification includes generic, but random sensitivities to travel time.} In line with \citet{cirillo2006evidence} and \citet{ory2004when}, we argue that individuals may exhibit both positive and negative valuations of travel time for the daily commute and therefore do not restrict the support of the respective heterogeneity distributions. 

We also explored several model specifications, in which different restrictions on the covariance matrix $\boldsymbol{\Sigma}$ of the multivariate normal mixing distribution were imposed. In the interest of parsimony and computational tractability, the final model specification only captures correlation in the sensitivities to travel time. 

Finally, we note that the final model specification includes intercepts for the travel options, whereby the option conventional car is treated as a reference. We allow for systematic heterogeneity in the mode-specific intercepts by parameterising them as functions of additional fixed parameters and socio-demographic attributes of the respondents. 

\subsection{Overall model evaluation}

To quantify the benefits of accommodating flexible substitution patterns and unobserved taste variation, we estimate several alternative models with varying levels of complexity. The first model (C-MNL) is a standard conditional multinomial logit model, which accommodates neither flexible substitution patterns nor unobserved taste variation. The second model (EC-MNL) is an error components multinomial logit model, which allows for flexible substitution patterns as outlined in Section \ref{S_modelling}. The third model (M-MNL I) is a mixed multinomial logit model, which additionally accounts for unobserved taste variation, whereby the off-diagonal elements of the covariance matrix of the multivariate normal mixing distribution are constrained to zero. The fourth model (M-MNL II) extends the third model by allowing for correlation between random taste parameters pertaining to travel times. Table \ref{table_modelComparison} provides a comparison of the estimated models. It can be seen that model M-MNL II yields the highest log-likelihood and $\rho^{2}$ values. Furthermore, the Bayesian Information Criterion \citep[BIC;][]{schwarz1978estimating} suggests that M-MNL II is the best of all considered models; likelihood-ratio tests indicate that M-MNL II provides a statistically significantly better fit than each of the competing models.

\begin{table}[h]
\centering
\small
\ra{1.2}
\begin{tabular}{@{}l S[table-format=5.2] S[table-format=5.2] S[table-format=5.2] S[table-format=5.2]  @{}}
\toprule
& 
\textbf{C-MNL} &
\textbf{EC-MNL} &
\textbf{M-MNL I} &
\textbf{M-MNL II}  \\
\midrule

No. of parameters &	29 &	34 &	40 &	43 \\
Log-likelihood & -6546.38 &	-5382.93 &	-5300.72 &	-5246.34 \\
$\rho^{2}$ &	0.09 &	0.25 &	0.26 &	0.27 \\
BIC & 13317.34	 & 11015.39 &	10900.88 &	10817.07\\
				
Likelihood-ratio test w.r.t. M-MNL II \\				
\quad $\chi^{2}$ & 2600.08 &	273.18 &	108.76 \\
\quad $df$ &	12 &	9 &	3 \\	
\quad $p$ &	\multicolumn{1}{r}{$< 0.001$} &	\multicolumn{1}{r}{$< 0.001$} &	\multicolumn{1}{r}{$< 0.001$} \\	
\midrule
\multicolumn{5}{l}{
\begin{minipage}[t]{0.8\textwidth}
\footnotesize 
Note: 
The null log-likelihood is $-7196.3$. 
BIC is computed as $\ln(N T) P - 2 LL$, where $NT = 512 \cdot 8$ is the sample size, $P$ is the number of model parameters and $LL$ is the log-likelihood at convergence.
\end{minipage}} \\
\bottomrule
\end{tabular}
\caption{Model comparison} \label{table_modelComparison}
\end{table}

\FloatBarrier
\subsection{Estimation results}

Table \ref{table_estimationRes} enumerates the estimation results for the models C-MNL, EC-MNL, M-MNL I and M-MNL II. In all models, the mean sensitivities to the attributes of the housing and travel options are statistically significantly different from zero and have the expected signs.\footnote{We re-iterate that the taste parameters pertaining to housing costs enter the utility function exponentially to facilitate the calculation of standard errors of willingness to pay measures \citep[see][]{carson2018new}. Under a conventional utility specification, the estimated cost sensitivities are negative and significantly different from zero.} In the interest of brevity, our subsequent discussion and analysis focus on the best-performing model M-MNL II.

First, we examine the estimation results for the taste parameters pertaining to housing attributes. The taste parameters pertaining to the attributes ``additional number of rooms'' and ``separate house vs. other dwelling types'' are subject to substantial random taste variation, as is indicated by the estimates of the scale parameters of the respective normal heterogeneity distributions. While the majority of the respondents prefer to live in a dwelling that is larger than their current home, a minority of the respondents either do not attend to the attribute in question or prefer to live in a smaller dwelling. Similarly, the majority of the respondents prefer to live in a separate house rather than in a town house or a unit, while a minority of the respondents either do not attend to the attribute or prefer to live in a town house or unit. We also observe that respondents generally prefer to live in a residential neighbourhood with mostly single family houses and in housing developments that are less than 15 years old. Furthermore, respondents generally prefer to live in close proximity to local services.

Second, we consider the estimation results for the taste parameters pertaining to the attributes of the travel options for the commute. The sensitivities to travel time are best interpreted, when they are denominated in currency units (see Section \ref{s_VOT}); nonetheless, we make the following observations at this point: The estimates of the mean parameters of the respective normal heterogeneity distributions are negative and significantly different from zero. Moreover, the estimates of the scale parameters of the heterogeneity distribution indicate considerable random taste variation in the sensitivities to travel time by the different travel options. It can further be seen that the majority of the estimates of the elements of the Cholesky factor of the covariance matrix capturing covariation in the sensitivities to travel time are significantly different from zero. We also observe that respondents are generally averse to congested travel conditions, while the sensitivities to the respective attribute are subject to considerable random taste variation. 

Third, we note that the estimation results for the mode-specific intercepts, which are specified as linear functions of socio-demographic attributes of the respondents, whereby the transport mode conventional car is treated as a reference option. The perceived benefit of autonomous driving is relatively greater among highly-educated respondents and current users of on-demand transportation services. 

Last, we observe that the estimates of the scale parameters of the normal error components. It can be seen that the estimates of all scale parameters are significantly different from zero. Overall, the estimation results corroborate the hypothesised nesting structure and substitution patterns. In addition, it can be seen that preferences for autonomous driving remain subject to substantial unobserved preference heterogeneity, even though systematic differences in preferences are systematically controlled for by parameterising the respective mode-specific intercept as a function of socio-demographic attributes of the decision-makers.

\begin{landscape}
\centering
\ra{1.2}
\footnotesize
\begin{longtable}{@{}
p{7cm}
S[table-format=1.4] S[table-format=1.4] c 
S[table-format=1.4] S[table-format=1.4] c 
S[table-format=1.4] S[table-format=1.4] c 
S[table-format=1.4] S[table-format=1.4]@{}}
\toprule
 & 
 \multicolumn{2}{c}{\textbf{MNL}} & \phantom{a} &
 \multicolumn{2}{c}{\textbf{EC-MNL}} & \phantom{a} &
 \multicolumn{2}{c}{\textbf{M-MNL I}} & \phantom{a} &
 \multicolumn{2}{c}{\textbf{M-MNL II}} \\
\cmidrule{2-3} \cmidrule{5-6} \cmidrule{8-9} \cmidrule{11-12}
\textbf{Attribute} & 
\multicolumn{1}{c}{\textbf{Est.}} & \multicolumn{1}{c}{\textbf{Std. err.}} & &
\multicolumn{1}{c}{\textbf{Est.}} & \multicolumn{1}{c}{\textbf{Std. err.}} & &
\multicolumn{1}{c}{\textbf{Est.}} & \multicolumn{1}{c}{\textbf{Std. err.}} & &
\multicolumn{1}{c}{\textbf{Est.}} & \multicolumn{1}{c}{\textbf{Std. err.}} 
\\
\midrule

\textbf{Housing attributes} \\
Housing cost [$-$AUD/week] divided by household income [AUD/week] times $10^{1}$ \\
\quad Owner---enters exponentially &
-0.6027$^{*}$ & 0.3446& & 
-0.5553$^{*}$ & 0.3320 & & 
-0.4104 & 0.3243 & & 
-0.3306 & 0.3124 \\
\quad Renter---enters exponentially &
-0.0937 & 0.1550 & & 
-0.0728 & 0.1524 & & 
-0.0027 & 0.1569 & & 
0.0190 & 0.1587 \\
Additional number of rooms \\
\quad Mean &
0.1532$^{***}$ & 0.0316 & &
0.1571$^{***}$ & 0.0321 & & 
0.1775$^{***}$ & 0.0405 & & 
0.1924$^{***}$ & 0.0433  \\
\quad Standard deviation &
& & & 
& & & 
0.3790$^{***}$ & 0.0724 & & 
0.4353$^{***}$ & 0.0742 \\
Separate house vs. other dwelling types  \\
\quad Mean &
0.5380$^{***}$ &  0.0456 & & 
0.5509$^{***}$ &  0.0463 & & 
0.6322$^{***}$ &  0.0673 & & 
0.6885$^{***}$ &  0.0720 \\
\quad Standard deviation &
& & & 
& & & 
0.9124$^{***}$ & 0.0921 & &
0.9829$^{***}$ & 0.0953 \\
Neighbourhood type is ``mostly single-family houses for residential use only'' &
0.1421$^{***}$ & 0.0511 & & 
0.1313$^{**}$ & 0.0518 & & 
0.1508$^{***}$ & 0.0573 & & 
0.1452$^{**}$ & 0.0597 \\
Age of housing development is 15 years or more &
-0.3639$^{***}$ & 0.0492 & &  
-0.3765$^{***}$ & 0.0500 & &  
-0.4147$^{***}$ & 0.0555 & &  
-0.4388$^{***}$ & 0.0581 \\
Local services available vs. not available &
0.4375$^{***}$ & 0.0538 & & 
0.4324$^{***}$ & 0.0546 & & 
0.5049$^{***}$ & 0.0610 & & 
0.5173$^{***}$ & 0.0633 \\
Walking time to local services is 10 minutes if local services are available &
0.2358$^{***}$ & 0.0574 & & 
0.2467$^{***}$ & 0.0583 & & 
0.2530$^{***}$ & 0.0643 & &
0.2652$^{***}$ & 0.0667 \\

\textbf{Commute attributes} \\

Travel cost [$-$ AUD]---enters exponentially&
-2.3581$^{***}$ & 0.1670 & &
-2.1346$^{***}$ & 0.1545 & &
-1.9947$^{***}$ & 0.1472 & &
-1.9562$^{***}$ & 0.1460 \\

Travel time by conventional car [h] \\
\quad Mean &
-1.8044$^{***}$ & 0.1994 & & 
-2.0752$^{***}$ & 0.2923 & & 
-3.4863$^{***}$ & 0.3979 & & 
-3.5725$^{***}$ & 0.4608 \\
\quad Standard deviation &
& & & 
& & & 
2.6849$^{***}$ & 0.3462 & & 
5.2152$^{***}$ & 0.3887$^{\#}$ \\
Travel time by self-driving car \\
\quad Mean &
-1.4537$^{***}$ & 0.2009 & & 
-2.2457$^{***}$ & 0.2952 & & 
-3.8282$^{***}$ & 0.3970 & & 
-3.3968$^{***}$ & 0.4233 \\
\quad Standard deviation & 
& & &
& & & 
2.4614$^{***}$& 0.2550 & & 
4.5621$^{***}$& 0.3644$^{\#}$ \\
Travel time by public transportation \\
\qquad Mean  & 
-1.2173$^{***}$ & 0.1935 & & 
-1.1489$^{***}$ & 0.2688 & & 
-2.5314$^{***}$ & 0.4316 & &
-2.6900$^{***}$ & 0.4306 \\
\qquad Standard deviation & 
& & & 
& & & 
3.0862$^{***}$ & 0.3318 & & 
4.6409$^{***}$ & 0.3620$^{\#}$ \\

Proportion of travel time spent in congestion [$-\%$]  \\
\quad Mean &
-0.5124$^{***}$ & 0.1000 & & 
-0.6951$^{***}$ & 0.1150 & & 
-0.8170$^{***}$ & 0.3026 & & 
-0.6971$^{**}$ & 0.2801 \\
\quad Standard deviation &
& & & 
& & & 
1.1555$^{***}$ & 0.1907 & & 
1.0206$^{***}$ & 0.1899 \\

Mode-specific intercepts (reference = conventional car) \\
\quad Self-driving car \\
\qquad Baseline &
-1.4494$^{***}$ & 0.1541 & & 
-1.8420$^{***}$ & 0.5271 & & 
-1.5286$^{***}$ & 0.5950 & & 
-2.3277$^{***}$ & 0.5058 \\
\qquad Female &
0.1248 & 0.0867 & &
0.2470 & 0.3388 & &
0.4404 & 0.3688 & &
0.1874 & 0.3441 \\
\qquad Age (reference = 30 to 49 years old)\\
\qquad \quad 18 to 29 years old &
0.1611 & 0.1227 & & 
0.1396 & 0.4844 & &  
-0.3850 & 0.5171 & &  
-0.0293 & 0.4679 \\
\qquad \quad 50 years old or older &
-0.4463$^{***}$ & 0.1119 & &
-0.7953$^{*}$ & 0.4319 & & 
-1.0492$^{**}$ & 0.4751 & &
-0.8021$^{*}$ & 0.4399 \\
\qquad Presence of children in household &
0.2015 & 0.0948 & &  
0.4369 & 0.3738 & &  
0.2960 & 0.4133 & &  
0.6163 & 0.3719 \\
\qquad Education level is Bachelors degree or higher &
0.6821$^{***}$ & 0.0919 & &  
1.2073$^{***}$ & 0.3657 & &  
1.3596$^{***}$ & 0.4329 & &  
1.3967$^{***}$ & 0.3693 \\
\qquad Ride-hailing service user  &
0.7549$^{***}$ & 0.0906 & & 
1.4067$^{***}$ & 0.3541 & & 
1.3254$^{***}$ & 0.3915 & & 
1.6003$^{***}$ & 0.3592 \\

\quad Public transportation \\
\qquad Baseline &
-1.2694$^{***}$ & 0.1454 & & 
-2.1577$^{***}$ & 0.5441 & & 
-1.7254$^{***}$ & 0.6596 & &
-2.6863$^{***}$ & 0.5840 \\
\qquad Female &
0.2120$^{***}$ & 0.0798 &&
0.4465 & 0.3642 && 
0.1105 & 0.7780 & & 
0.3471 & 0.3748 \\
\qquad Age (reference = 30 to 49 years old)\\
\qquad \quad 18 to 29 years old &
-0.2298$^{*}$ & 0.1214 & &  
-0.4000 & 0.5244 & &  
-0.6607 & 0.5580 & & 
-0.3639 & 0.5116 \\
\qquad \quad 50 years old or older &
-0.2102$^{**}$ & 0.0966 & & 
-0.4670 & 0.4601 & & 
-0.8040 & 0.5003 & & 
-0.3207 & 0.4666 \\
\qquad Presence of children in household &
-0.3766$^{***}$ & 0.0909 & &  
-0.7289$^{*}$ & 0.4073 & &  
-0.7796$^{**}$ & 0.4411 & &  
-0.5035 & 0.4510 \\
\qquad Education level is Bachelors degree or higher &
0.7590$^{***}$ & 0.0842 & &  
1.6376$^{***}$ & 0.3924 & & 
1.6405$^{***}$ & 0.4525 & &  
1.6714$^{***}$ & 0.3926 \\
\qquad Ride-hailing service user  &
0.4708$^{***}$ & 0.0847 & & 
1.0211$^{***}$ & 0.3842 & & 
0.8472$^{***}$ & 0.4164 & & 
1.3190$^{***}$ & 0.3995 \\

\textbf{Cholesky} \\
Travel time by conventional car [0.25 h] vs. \\
\quad Travel time by conventional car [0.25 h] &
& & & 
& & & 
& & & 
1.3038$^{***}$ & 0.0984 \\
\quad Travel time by self-driving car [0.25 h] &
& & & 
& & & 
& & & 
1.1209$^{***}$ & 0.0933 \\
\quad Travel time by public transit car [0.25 h] &
& & & 
& & & 
& & & 
1.0202$^{***}$ & 0.1038 \\

Travel time by self-driving car [0.25 h] vs. \\
\quad Travel time by self-driving car [0.25 h] &
& & & 
& & & 
& & & 
0.2105$^{***}$ & 0.0619 \\
\quad Travel time by public transit car [0.25 h] &
& & & 
& & & 
& & & 
-0.1358 & 0.0911 \\

Travel time by public transit car [0.25 h] vs. \\
\quad Travel time by public transit car [0.25 h] &
& & & 
& & & 
& & & 
0.5356$^{***}$ & 0.0956 \\

\textbf{Scales of heteroskedastic error components} \\
Conventional car &
& & &
2.6001$^{***}$ & 0.2157 & &
2.2939$^{***}$ & 0.2597 & &
2.7307$^{***}$ & 0.1989 \\
Self-driving car &
& & &
1.5594$^{***}$ & 0.2589 & &
0.5127$^{*}$ & 0.2832 & &
1.2486$^{**}$ & 0.2798 \\
Public transportation &
& & &
2.2602$^{***}$ & 0.2034 & & 
0.7268$^{*}$ & 0.4313 & &
1.5546$^{***}$ & 0.4696 \\

\midrule
\multicolumn{12}{l}{
\begin{minipage}[t]{1.3\textwidth}
\footnotesize 
Note: 
$^{*}$ $\mbox{p-value} \in (0.05, 0.1]$, $^{**}$ $\mbox{p-value} \in (0.01, 0.05]$, $^{***}$ $\mbox{p-value} \leq 0.01$;
$^{\#}$ computed via the method proposed by \citet{krinsky1986approximating}.
\end{minipage}} \\
\bottomrule
\caption{Estimation results} \label{table_estimationRes}
\end{longtable}
\end{landscape}

\subsection{Value of time} \label{s_VOT}

The value of time (VOT), i.e. the marginal rate of substitution of travel time and cost, is a central quantity in transport planning \citep{small2012valuation}. In the present application, VOTs for commuting by conventional car, self-driving car and public transit can be defined in terms of either housing or travel costs. In either case, VOT distributions are given by the ratio of the random travel time sensitivities and the fixed cost parameter. Since the random travel time sensitivities are normally distributed, the VOT distributions are also normal. 
Table \ref{table_VOT} enumerates the point estimates and confidence intervals of the parameters of the VOT distributions for models M-MNL I and II. In what follows, we discuss the results of model M-MNL II in more detail. 

First, we examine the VOT distributions that are defined in terms of travel costs. The mean VOTs fall within the range from 19.0 AUD/h to 25.3 AUD/h: The mean VOT is greatest for commuting by conventional car  (25.3 AUD/h), followed by commuting by self-driving car (24.0 AUD/h) and is smallest for commuting by public transit (19.0 AUD/h). However, the differences in the point estimates are not statistically significant, because the confidence intervals of the estimates overlap. 
In comparison with the literature and government guidelines, the estimated mean values of commute travel time savings appear plausible: Relying on stated preference data collected in Brisbane, Australia in 2008, \citet{hensher2011embedding} suggest a mean value of commute travel time savings for travel by conventional car of 17.60 AUD/h. Using data sourced from the Sydney Household Travel Surveys of the years 2007/08, 2008/09 and 2009/10, \citet{ho2013tour} estimate that the values of in-vehicle, wait and walk times for weekday travel are 6.81 AUD/h, 11.28 AUD/h and respectively 14.68 AUD/h (denominated in 2008 AUD). Drawing from stated preference data collected in Sydney in 2014, \citet{ho2016vehicle} determine mean VOTs for work travel of 13.65 AUD/h for car drivers accompanied by another passenger and of 8.30 AUD/h for unaccompanied car drivers. Australian government guidelines for cost-benefit-analyses recommend that travel time by conventional car shall be valued at 14.99 AUD/h for private travel and at 48.63 AUD/h for business travel; while travel time of bus passengers shall be valued at 14.99 AUD/h \citep{transport2018australian}. 

Second, we examine the estimated distributions of the values of commute travel time savings in terms of housing costs. Recall that the sensitivities to housing costs vary systematically in tenure type and household income. Due to the cost over income specification of the cost attribute, the sensitivities to housing costs are given by the ratio of the estimate of the appropriate fixed cost parameter and household income. In Table \ref{table_VOT}, we report the mean and standard deviations of the values of commute travel times at the mean income levels of owners and renters. The point estimates suggest that on average, owners value commute travel time savings higher than renters. However, overlapping confidence intervals suggest that differences in mean VOT across owners and renters are not statistically significantly different from zero.

Third, we compare the VOT estimates in terms of travel and housing costs. Note that VOT in terms of travel costs is reported in AUD/h, whereas VOT in terms of housing costs is reported in 10 AUD/week/h. The VOT estimates in terms of the two cost types can be meaningfully compared, if VOT in terms of travel costs is scaled to a weekly basis. To that end, we suppose a weekly commute travel time budget of ten hours and multiply the reported parameter estimates and confidence interval bounds for VOT in terms of travel and housing costs by 10 h and 1 h, respectively. Under this assumption, it can be seen that on average, respondents value commute travel time savings more in terms of housing costs than in terms of travel costs in the majority of cases. For example, the mean VOT for commuting by self-driving car in terms of travel cost is 240.2 AUD/week, whereas the mean VOT for commuting by self-driving car in terms of housing costs is 1061.3 AUD/week for owners and 519.5 for AUD/week for renters. For conventional car and self-driving car, the differences in mean VOT are statistically significant, as the corresponding confidence interval do not overlap. For public transit, the confidence intervals do not overlap for owners but overlap to a small extent for renters. 

By and large, the same substantive insights can be derived from model M-MNL I, which in contrast to model M-MNL II does not capture covariation in sensitivities to travel time. From Table \ref{table_VOT}, it can be seen that the point estimates of the mean parameters of the VOT distributions and the corresponding confidence intervals match closely. Like model M-MNL II, model M-MNL I does not suggest that the differences in the mean VOT are statistically significant. 

\begin{landscape}
\small
\begin{table}[h]
\centering
\small
\ra{1.2}
\begin{tabular}{@{}
p{7cm}
S[table-format=3.2] c c 
S[table-format=2.2] c c
S[table-format=3.2] c c 
S[table-format=3.2] c @{}}
\toprule

& \multicolumn{5}{c}{\textbf{M-MMNL I}} & & \multicolumn{5}{c}{\textbf{M-MMNL II}} \\
\cmidrule{2-6} \cmidrule{8-12}
 & 
 \multicolumn{2}{c}{\textbf{Mean}} & &
 \multicolumn{2}{c}{\textbf{Std. dev.}} & &
 \multicolumn{2}{c}{\textbf{Mean}} & &
 \multicolumn{2}{c}{\textbf{Std. dev.}} \\
\cmidrule{2-3} \cmidrule{5-6} \cmidrule{8-9} \cmidrule{11-12}
\textbf{Value of travel time} & 
\multicolumn{1}{c}{\textbf{Est.}} & \multicolumn{1}{c}{\textbf{95\%-CI}} & &
\multicolumn{1}{c}{\textbf{Est.}} & \multicolumn{1}{c}{\textbf{95\%-CI}} & &
\multicolumn{1}{c}{\textbf{Est.}} & \multicolumn{1}{c}{\textbf{95\%-CI}} & &
\multicolumn{1}{c}{\textbf{Est.}} & \multicolumn{1}{c}{\textbf{95\%-CI}} 
\\
\midrule

In terms of terms of travel cost [AUD/h] \\
\quad Conventional car &
25.62 & $[16.53, 38.42]$ & & 
19.73 & $[13.21, 28.29]$ & &
25.26 & $[15.79, 38.67]$ & & 
36.88 & $[26.85,  50.06]$\\
\quad Self-driving car &
28.14 & $[18.34, 41.90]$ & & 
18.09 & $[12.58, 25.43]$ & &
24.02 & $[15.05, 36.36]$ & & 
32.26 & $[23.35,   44.01]$\\
\quad Public transit &
21.57 & $[12.96, 33.62]$ & & 
18.61 & $[12.47, 26.92]$ & &
19.02 & $[10.89, 30.16]$ & & 
32.82 & $[24.06, 45.05]$\\

In terms of housing cost [10 AUD/week/h] \\
\quad Owner (household income = 2244.7 AUD/week)& \\
\qquad Conventional car &
117.97 & $[59.00, 230.17]$ & & 
90.85 & $[44.86, 178.77]$ & &
111.62 & $[56.84, 212.77]$ & & 
162.94 & $[87.69, 302.60]$\\
\qquad Self-driving car &
129.54 & $[65.63, 246.97]$ & & 
83.29 & $[42.78, 160.91]$ & &
106.13 & $[54.30, 201.82]$ & & 
142.54 & $[76.30, 268.66]$\\
\qquad Public transit &
99.30 & $[47.95, 198.02]$ & & 
85.66 & $[43.07, 168.58]$ & &
84.05 & $[40.82, 164.21]$ & & 
145.00 & $[78.69, 272.64]$ \\
\quad Renter (household income = 1558.7 AUD/week) & \\
\qquad Conventional car &
54.49 & $[40.97, 159.82]$ & & 
41.96 & $[31.15, 124.13]$ & &
54.63 & $[39.47, 147.74]$ & & 
79.75 & $[60.89, 210.11]$ \\
\qquad Self-driving car &
59.83 & $[45.57, 171.48]$ & & 
38.47 & $[29.70, 111.73]$ & &
51.95 & $[37.70, 140.14]$ & & 
69.77 & $[52.98, 186.55]$ \\
\qquad Public transit &
45.86 & $[33.30, 137.49]$ & & 
39.56 & $[29.91, 117.06]$ & &
41.13 & $[28.35, 114.02]$ & &
70.97 & $[54.64, 189.31]$\\

\midrule
\multicolumn{6}{l}{
\begin{minipage}[t]{0.8\textwidth}
\footnotesize 
Note: 
95\%-CI = 95\%-confidence intervals computed via the \citet{krinsky1986approximating} method. 
\end{minipage}} \\
\bottomrule
\end{tabular}
\caption{Estimated distributions of the value of time for commuting by conventional car, self-driving car and public transit in terms of travel cost and housing cost for models M-MNL I and M-MNL II} \label{table_VOT}
\end{table}
\end{landscape}

\FloatBarrier
\section{Discussion and conclusions} \label{S_conclusion}

In this paper, we make an effort to empirically quantify the potential impacts of autonomous driving on travel behaviour and land use by investigating stated preferences for combinations of residential locations and travel options for the commute in the context of autonomous driving. Our analysis draws from a stated preference survey, which was completed by 512 residents from the Sydney metropolitan area in Australia, and provides insights into travel time valuations for autonomous driving in a long-term decision-making context. For the inferential analysis of the stated choice data, several mixed logit models were estimated. Overall, the empirical results suggest that no changes in the valuation of travel time due to the advent of AVs should be expected. 

Several simulation-based studies investigating the impacts of autonomous driving on travel behaviour and residential location choices assume that VOT for travel by AV can be discounted at rates of up to 100\% relative to VOT for travel by conventional car \citep[see][]{gelauff2017spatial, thakur2016urban, zhang2018residential}. It is surmised that VOT for travel by AV is lower than VOT for travel by conventional car, because AVs may render car travel less burdensome, in part by enabling productive time use during travel. In contrast, other studies have questioned whether the advent of AVs will in fact lead to substantive reductions in VOT. For example, \citet{singleton2019discussing} contends that reductions in VOT may more likely stem from increases in subjective well-being due to decreased driving stresses rather than from the possibility to make productive use of travel time. Referencing \citet{le_vine2015autonomous}, \citet{singleton2019discussing} suggests that the level of ride comfort of AVs may be insufficient for extensive travel-based multitasking. Moreover, \citet{singleton2019discussing} asserts that in general, travel-based multitasking is mostly driven by the desire to pass time rather than by higher-order goals for productive and efficient time use. Relatedly, \citet{pudane2018will} suggest that substantial changes in individuals' activity scheduling preferences are necessary, before the possibility of making productive use of travel time on AVs can affect VOT. 

Overall, the findings of this study provide support to the view that the immediate and instantaneous potential of AVs to substantively reduce VOT and to thus affect residential location preferences may be limited \citep[also see in particular][]{singleton2019discussing}. One immediate implication of this insight is that caution should be exercised in making assumptions about changes in VOT in simulation-based analyses investigating the impacts of autonomous driving on travel behaviour and residential location choices \citep[also see][]{soteropoulos2018impacts}. Nonetheless, the findings of our study also suggest that sensitivities to travel time by AV are subject to considerable random taste variation, which in turn indicates substantial heterogeneity in the perception of the benefits of AVs in terms of their potential for reducing driving stresses and for enabling productive time use. Investigating the sources of this heterogeneity is an intriguing avenue for future research.

In addition, there are several other directions in which future research may build on the work presented in the current paper. First and foremost, the current study solely relies on stated preference data. Even though stated preference studies are pivotal in capturing the relative importance of attributes of new product and services, it should be acknowledged that the estimates derived in the current study are likely subject to a hypothetical bias and may therefore not accurately reflect individuals' preferences, when AVs actually become widely available, as it is difficult for respondents to imagine the use of AVs \citep[also see][]{beck2016can, krueger2016preferences, milakis2017policy}. Current preferences for travel by AVs may be dominated by aversions to and fears of the unknown technology as well as by uncertainty about the value of the new technology. One remedy to this issue is to triangulate stated choice techniques with other methods that allow respondents to immerse into a virtual reality of autonomous driving such as well-designed driving simulator studies or naturalistic experiments \citep[see e.g.][]{harb2018projecting}. Second, the stated choice experiment presented as part of the current study does not explicitly specify the business models under which AVs may become available. It has been argued that the autonomous driving technology may be well suited for the realisation of autonomous on-demand transportation services \citep{burns2013vision, fagnant2014travel, krueger2016preferences}. The availability and pricing structures of such services may likely affect preferences for travel by AVs as well as residential location preferences. Third, it has been shown that VOT estimates may differ depending on whether the preference elicitation occurs in short- or in a long-term decision-making context \citep{beck2017valuing}. Along these lines, it may be useful to compare how preferences for autonomous driving vary across different elicitation contexts. Finally, residential location choices are typically made at the household-level and are therefore subject to group interactions. To enhance the realism of stated preference studies investigating the potential impacts of AVs on residential location preferences, it may thus be useful to elicit preferences in a group-based decision-making context. 

\section*{Acknowledgements}

We would like to thank two anonymous reviewers for their critical assessment of our work. A preliminary version of this paper was presented at the 98\textsuperscript{th} Annual Meeting of the Transportation Research Board; we would like to thank the anonymous reviewers of this preliminary version for their constructive comments. The authors acknowledge support from the Australian Research Council (LP160100450).

\section*{Author contribution statement}

RK: conception and design, data collection, preparation and analysis, manuscript writing and editing. \\
THR: conception and design, data collection and analysis, manuscript editing, supervision. \\
VVD: data collection.

\newpage
\bibliographystyle{apalike}
\bibliography{2016_pva.bib}

\newpage
\appendix
\section{Description of the stated choice data}

\begin{table}[h]
\small
\centering
\ra{1.2}
\begin{tabular}{@{}p{5cm} S[table-format=3.1] S[table-format=3.1] S[table-format=3.2] S[table-format=4.1] S[table-format=3.2] S[table-format=4.0]@{}}
\toprule
\textbf{Attribute}  & \multicolumn{1}{c}{\textbf{Min.}} & \multicolumn{1}{c}{\textbf{Median}} & \multicolumn{1}{c}{\textbf{Mean}} & \multicolumn{1}{c}{\textbf{Max.}}  & \multicolumn{1}{c}{\textbf{Std. dev.}} & \multicolumn{1}{c}{\textbf{N}} \\
\midrule
Mortgage payment [AUD/week]      &	143 & 475 & 537.4 & 4200 & 431.4 & 2256\\

Rent [AUD/week]      &	143 & 405 & 438.8 & 1281 & 191.5 & 1840\\

Additional number of rooms & 0 & 1 & 1.0 & 2 & 0.8 & 4096\\

Dwelling type (dummies) \\
\quad Unit			& 0 & 0 & 0.26 & 1 & 0.44 & 1074 \\
\quad Townhouse	& 0 & 0 & 0.36 & 1 & 0.48 & 1472 \\
\quad Separate house & 0 & 0 & 0.38 & 1 & 0.49 & 1550 \\

Neighbourhood type (dummies) \\
\quad Mostly single-family houses for residential use only		& 0 & 0 & 0.36 & 1 & 0.48 & 1454 \\
\quad Mix of low-rise and multi-storey buildings for mixed residential and commercial use	& 0 & 0 & 0.33 & 1 & 0.47 & 1332 \\
\quad High-rise buildings for mixed residential and commercial use & 0 & 0 & 0.32 & 1 & 0.47 & 1310 \\

Age of development (dummies) \\
\quad 0--5 years		& 0 & 0 & 0.38 & 1 & 0.49 & 1552 \\
\quad 5--15 years	& 0 & 0 & 0.33 & 1 & 0.47 & 1362 \\
\quad More than 15 years & 0 & 0 & 0.29 & 1 & 0.45 & 1182 \\

Local services (dummies) \\
\quad No shops		& 0 & 0 & 0.27 & 1 & 0.44 & 1103 \\
\quad Basic shops and restaurants	& 0 & 0 & 0.34 & 1 & 0.47 & 1388 \\
\quad Basic plus specialty shops and restaurants & 0 & 0 & 0.39 & 1 & 0.49 & 1605 \\

Walking time to local services, if local services are available [minutes] & 10 & 20 & 19.4 & 30 & 8.1 & 2993 \\

\bottomrule
\end{tabular}
\caption{Summary statistics of the attribute levels of chosen alternatives for the housing option (N = 4,096)} \label{table_dataChosenHousing}
\end{table}

\begin{table}[h]
\small
\centering
\ra{1.2}
\begin{tabular}{@{}p{5cm} S[table-format=2.1] S[table-format=2.1] S[table-format=2.1] S[table-format=3.1] S[table-format=2.1] S[table-format=4.0]@{}}
\toprule
\textbf{Attribute}  & \multicolumn{1}{c}{\textbf{Min.}} & \multicolumn{1}{c}{\textbf{Median}} & \multicolumn{1}{c}{\textbf{Mean}} & \multicolumn{1}{c}{\textbf{Max.}}  & \multicolumn{1}{c}{\textbf{Std. dev.}} & \multicolumn{1}{c}{\textbf{N}} \\
\midrule
\textbf{Conventional car} \\
Travel cost [AUD]      &	3.2 & 5.0 & 6.0 & 18.0 & 2.8 & 1452\\
Travel time [minutes] &	32 & 42 & 48.9 & 120 & 20.5 & 1452\\
Proportion of travel time spent in congested traffic conditions [\%] &	10 & 35 & 35.6 & 65 & 22.9 & 1452\\

\textbf{Self-driving car} \\
Travel cost [AUD]      &	3.2 & 5.6 & 6.3 & 18.0 & 2.8 & 1116\\
Travel time [minutes] &	32 & 45 & 51.6 & 120 & 21.0 & 1116\\
Proportion of travel time spent in congested traffic conditions [\%] &	10 & 35 & 33.2 & 65 & 21.9 & 1116\\

\textbf{Public transportation} \\
Travel cost [AUD]      &	3.2 & 5.6 & 6.5 & 18.0 & 2.9 & 1528\\
Travel time [minutes] &	32 & 45 & 52.9 & 120 & 21.3 & 1528\\

\bottomrule
\end{tabular}
\caption{Summary statistics of the attribute levels of chosen alternatives for the travel options for the commute (N = 4,096)} \label{table_dataChosenMob}
\end{table}

\end{document}